\documentclass{ldr-article}

\usepackage{booktabs}
\usepackage{hyperref}
\hypersetup{
    colorlinks=true,
    linkcolor=blue,
    filecolor=magenta,      
    urlcolor=cyan,
    }
\usepackage{enumitem}
\usepackage[para]{threeparttable}

\usepackage{tikz}
\usepackage{tikz-qtree}
\usetikzlibrary{trees}	
\usepackage{subfigure}
\usepackage{multirow}

\usepackage{cite}
\title{\color{black} Synthetic Data in Radiological Imaging: \\Current State and Future Outlook}

\author{\textit{E. Sizikova\footnote{Corresponding author email: elena.sizikova@fda.hhs.gov }}, \textit{A. Badal}, \textit{J. G. Delfino}, \textit{M. Lago}, \textit{B. Nelson}, \textit{N. Saharkhiz}, \\ 
\textit{B. Sahiner}, \textit{G. Zamzmi}, \textit{A. Badano} \\ \\ 
Office of Science and Engineering Laboratories\\  
Center for Devices and Radiological Health\\ 
U.S. Food and Drug Administration\\ 
Silver Spring, MD 20993 USA}

\begin{document}

    \maketitle

    \begin{abstract}
 A key challenge for the development and deployment of artificial intelligence (AI) solutions in radiology is solving the associated data limitations. Obtaining sufficient and representative patient datasets with appropriate annotations may be burdensome due to high acquisition cost, safety limitations, patient privacy restrictions or low disease prevalence rates. In silico data offers a number of potential advantages to patient data, such as diminished patient harm, reduced cost, simplified data acquisition, scalability, improved quality assurance testing, and a mitigation approach to data imbalances. We summarize key research trends and practical uses for synthetically generated data for radiological applications of AI. Specifically, we discuss different types of techniques for generating synthetic examples, their main application areas, and related quality control assessment issues.  We also discuss current approaches for evaluating synthetic imaging data. Overall, synthetic data holds great promise in addressing current data availability gaps, but additional work is needed before its full potential is realized. 
    \end{abstract}
    


\section{Introduction}

Artificial Intelligence (AI) applications are becoming more and more prevalent in radiology and other types of medical imaging applications. AI techniques are used to aid clinical professionals in faster and more accurate detection of findings, {\color{black} optimize image quality} while reducing dose, and improve other facets of analyzing complex and multidimensional radiological data. A key feature of AI is its reliance on large-scale datasets for learning meaningful features. The goal of this paper is to review and discuss the emerging use of synthetic data for AI applications in radiology. 

\begin{table}[t!]
    \centering
    \resizebox{1.0\textwidth}{!}{
    \begin{threeparttable}
    \begin{tabular}{|l|c|c|c|c|c|}
        \hline
        {\textbf{Dataset}} & {\textbf{Anatomy}} & {\textbf{Imaging Modality}} & {\textbf{Data Generation Technique}} & {\textbf{Task}} \\
        \hline
        pGAN~\cite{sun2021deep}\tnote{a} & Vertebral units & MRI & GAN & Unit classification  \\
        \hline
        SinGAN-Seg~\cite{thambawita2022singan}\tnote{a} & Colon  & Endoscopy & GAN &  Polyp segmentation  \\
        \hline
        COVID-19 chest X-rays~\cite{zunair2021synthesis}\tnote{a} & Chest  & X-ray & GAN & Image classification   \\
        \hline
        Med-DDPM~\cite{dorjsembe2024conditional} & Brain & MRI & DDPM &  Brain tumor segmentation \\
        \hline
        Echo from noise~\cite{stojanovski2023echo}\tnote{a} & Heart  & Ultrasound & DDPM & Cardiac segmentation   \\
        \hline
        Awesome Lungs~\cite{ali2022spot}\tnote{a} & Lung  & CT, X-Ray & DDPM & Lung disease diagnosis \\
        \hline
        {\color{black} Synthetic CSAW 100k Mammograms}~\cite{pinaya2023generative}\tnote{a} & Breast & DM & DDPM & Classification masking of cancer \\
        \hline
        Sarno~\cite{Sarno2021_digitaldataset}\tnote{a} & Breast & CT, DM, DBT  & Physical modeling & Development of a platform for virtual clinical trials\\
        \hline
        VICTRE~\cite{Badano2018victre}\tnote{a} & Breast  & DM, DBT & Physical modeling &  Lesion detection  \\
        \hline
        M-SYNTH~\cite{sizikova2023knowledge}\tnote{a} & Breast  & DM & Physical modeling &  Lesion detection \\
        \hline
        Synthetic renal ASL data~\cite{brumer2022synthetic}\tnote{a} & Kidney  & MRI & Physical modeling &  Registration, quantification, segmentation  
\\
        \hline
    \end{tabular}
        \begin{tablenotes}
        \footnotesize
        \item[a] Publicly available dataset, according to the publisher
    \end{tablenotes}
    \end{threeparttable}
    }
    \caption{List of available synthetic radiologic datasets.}
    \label{tab:synthetic_datasets}
\end{table}

AI applications are often reliant on neural networks to perform predictions such as classification, segmentation or detection of objects of interest. Neural networks require large and diverse data collections to perform appropriate training and evaluation procedures. However, collecting sufficient examples from real patient sources comes with limitations due to patient privacy concerns, acquisition and annotation difficulties, high cost, and other challenges common to obtaining and sharing medical imaging datasets. Synthetically generated radiological data has been proposed to address some of these challenges and has become increasingly popular and realistic. Such data has been explored across different research domains, but the lack of consistent terminology has prevented a more unified and systematic study of synthetic data for radiology. For instance, the AI community has widely explored generative model techniques for the generation of synthetic data~\cite{kazerouni2023diffusion,singh2021medical}, while the biomedical and clinical community have studied digital twin~\cite{pesapane2022digital} and in silico medicine applications~\cite{Badano2018victre}. A summary of existing synthetic radiological datasets can be found in Table~\ref{tab:synthetic_datasets}. The potential benefits of the use of synthetic radiological data are:
\begin{itemize}[leftmargin=*,itemsep=0pt,topsep=0pt]
\item \textbf{Reduced patient harm or risk.} Patient medical procedures (including imaging acquisition data collection) comes with inherent {\color{black} health risks} (e.g., radiation exposure) or privacy risks (e.g., leak of Personal Identifiable Information (PII)), while the use of synthetic data typically relieves these risks.
\item \textbf{Reduced time, lowered cost, and simplified acquisition. } Synthetic examples are often much easier and cost efficient to  acquire, preprocess, store and maintain than patient data.
\item \textbf{Scalability.} Large volumes of synthetic examples can be generated on demand when larger sample sizes are needed.
\item \textbf{Quality assurance.} Synthetic examples can serve as a test-bed for evaluating AI algorithms to rapidly test comparative trends, or potentially support safety and effectiveness evaluations.
\item \textbf{Mitigation of data imbalances.} Synthetic examples can be conditionally generated according to manually specified distributions, addressing known class imbalances. The resulting process may address imbalances and minimize bias, increasing diversity and enriching patient datasets. Conditional generation allows creating rare cases or under-represented populations. 
\end{itemize}

\begin{figure}[t]
    \centering    
    \subfigure[{\scriptsize $4.44\times 10^9$}]{\includegraphics[width=2.0cm]{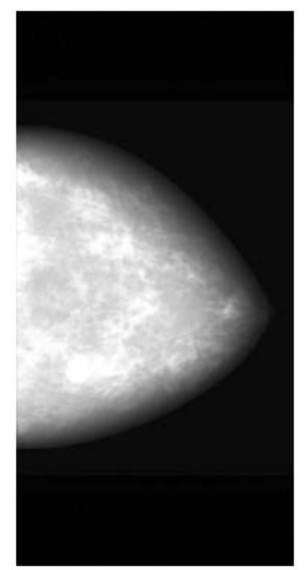}}
    \subfigure[{\scriptsize $8.88\times 10^9$}]{\includegraphics[width=2.0cm]{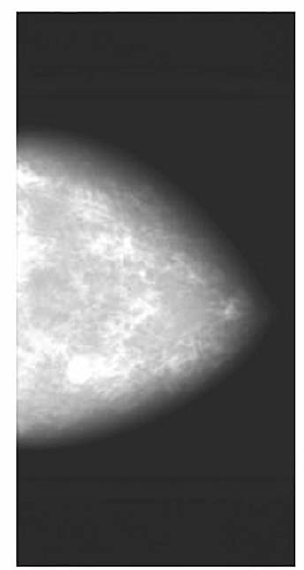}}
    \subfigure[{\scriptsize $1.33\times 10^{10}$}]{\includegraphics[width=2.0cm]{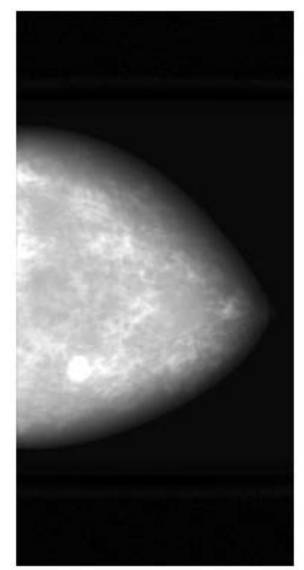}}
    \subfigure[{\scriptsize $1.78\times 10^{10}$}]{\includegraphics[width=2.0cm]{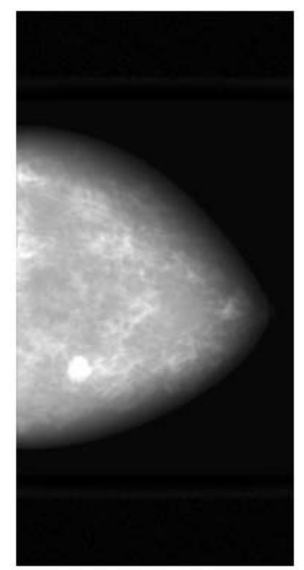}}
    \subfigure[{\scriptsize $2.22\times 10^{10}$}]{\includegraphics[width=2.0cm]{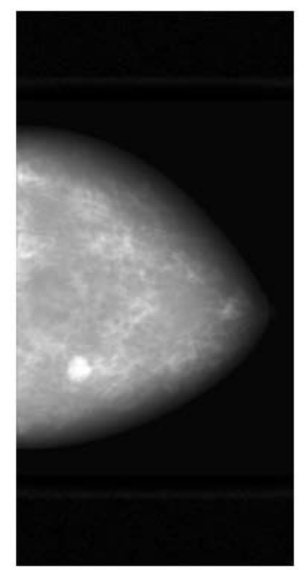}}
    \caption{Properties of the digital object and acquisition system models can be controlled during synthetic data generation process. Shown is the variation in imaging dose (number of Monte Carlo histories) generated with the VICTRE pipeline for digital mammography simulation~\cite{Badano2018victre} for a digital breast model~\cite{graffNewOpensourceMultimodality2016} with fatty breast density and mass model~\cite{de2015computational} with 5~mm radius (adapted from \cite{sizikova2023knowledge}).}
\label{fig:imaging_variations}
\end{figure}

 To date, synthetic data use has been explored in a variety of radiological AI uses. First, and most commonly, synthetic examples have been used as a source of rich and annotated training datasets, either on their own or combined with real patient examples to train an AI model, filling in gaps in data availability, particularly for underrepresented subgroups. Second, synthetic data can be used for generating standardized testing examples that would be otherwise too difficult to acquire from real patients. One notable emerging application is the use of synthetic data for in silico clinical trials. To accomplish these goals, synthetic data generators are becoming increasingly realistic and can be tuned to mimic properties of a radiologic device and the human anatomy (see Figure~~\ref{fig:imaging_variations} for an example).

The limitations of synthetic data depend on how synthetic examples were generated, how realistic they are, and whether their use has more benefits than shortcomings. We argue that this choice is application dependent. Synthetic data applications in radiology have been explored by the biomedical engineering, AI and clinical communities, but the differences in terminology and disconnect have prevented a more unified integration. To address existing gaps, we summarize the key uses of synthetic data for AI across radiological modalities to identify current and future trends in both ongoing research and current applications.

\section{Terminology}
A general definition of synthetic data in health care has been proposed as artificial data that mimic the properties and relationships seen in real {\color{black} patient data~\cite{myles2023potential}}. Synthetic examples are examples that have been partially or fully generated using computational techniques rather than acquired from a human subject by a physical system. The techniques used to generate synthetic examples (images and objects), described later in this article, vary in the fundamental origin of the information and are typically either knowledge-based or image-based approaches. 

The terms in silico imaging and in silico trials are closely related concepts which encompass computational approaches for generating data and evaluating imaging technology using computational models. \textit{In silico medicine} refers to the discipline that encompasses the use of patient-specific computer simulations involving all aspects of the prevention, diagnosis,
prognostic assessment, and treatment of disease~\cite{viceconti2016silico}. In turn, as defined in \cite{badano2023stochastic}, \textit{in silico imaging trials} are ``computational studies that seek to ascertain the performance of a medical device for the intended population, collecting this information entirely in the digital world via computer simulations''. Badano et al. \cite{badano2023stochastic} discusses techniques for generating digital cohorts, referring to groups of digital stochastic human models that share common characteristics and are selected for participation within in silico trials.

\begin{table}[]
\resizebox{1.0\textwidth}{!}{
{\color{black}
\begin{tabular}{|l|l|l|l|l|}
\hline
\textbf{Manuscript} & \textbf{Model Type} & \begin{tabular}[c]{@{}l@{}}\textbf{Imaging}\\ \textbf{Modality}\end{tabular} & \begin{tabular}[c]{@{}l@{}}\textbf{Anatomic}\\ \textbf{Site}\end{tabular} & \begin{tabular}[c]{@{}l@{}}\textbf{Comments}\end{tabular} \\ \hline
\multicolumn{5}{|c|}{}  \\ 
\multicolumn{5}{|c|}{\textbf{Techniques for Synthetic Data Generation}}  \\ 
\multicolumn{5}{|c|}{}  \\ 
\hline

Goodfellow~\cite{goodfellow2020generative} &     GAN    & Multiple   &  Multiple    &  Original GAN implementation                   \\ \hline

Kossen~\cite{kossen2022toward}             &     GAN    & MRI   &   Brain      &    Differentially private generation of image patches for brain vessel segmentation          \\ \hline

Jiang~\cite{jiang2018tumor}                      &     GAN    & CT    &  Lung    &    Adversarial domain adaptation (CT to MRI) for improved tumor segmentation  \\ \hline 

Xia~\cite{xia2020pseudo}                       &     GAN    & MRI    & Brain       &    Creation of patient-specific healthy examples from given pathological ones                         \\ \hline

Zhu~\cite{zhu2017unpaired}          &     GAN    &  Any  & Any  &   CycleGAN: unpaired image-to-image translation                              \\ \hline

Bora~\cite{bora2018ambientgan}      &     GAN    &  Any  &  Any   &   AmbientGAN: generative learning approach from noisy inputs                            \\ \hline

Zhou~\cite{zhou2022learning}       &     GAN    &  Multiple  & MRI     &    Generative learning on noisy, multi-resolution inputs applied to brain and knee data                       \\ \hline

Liu~\cite{liu2022free}            &     GAN    &  CT  &  Liver        &    Synthetic lesion synthesis for improved tumor segmentation                       \\ \hline

Jiang~\cite{jiang2020covid}       &     GAN    &  CT  &   Chest and Torso     &    COVID-19 CT synthesis using conditional generative learning                      \\ \hline

Kobyzev~\cite{kobyzev2020normalizing}               &     NF    &  Any  & Any  & Original Normalizing Flows (NF) implementation                                 \\ \hline

Denker~\cite{denker2020conditional}               &     NF    &  CT  & Chest and Torso  & Conditional NF for low-dose CT reconstruction                                 \\ \hline

Hajij~\cite{hajij2022normalizing}               &     NF    &  Multiple  & Multiple  & Comparison of NF to other generative models in medical image generation \\ \hline

Kingma~\cite{kingma2013auto}               &     VAE    &  Any  & Any  & Original VAE (VAE) implementation \\ \hline

Ahmad~\cite{ahmad2022brain}               &     VAE and GAN    &  MRI  & Brain  & VAE-GAN implementation for brain tumor MRI generation to avoid mode collapse \\ \hline

Cui~\cite{cui2022pet}               &     VAE    &  PET  & Chest and Torso  & Denoising and uncertainty estimation for PET \\ \hline
Ho~\cite{ho2020denoising}         &     DDPM    & Any   &  Any    &    Original diffusion model implementation                            \\ \hline

Khosravi~\cite{khosravi2023few}               &     DDPM    &  X-ray  & Pelvis  &    Diffusion models for few-shot segmentation                              \\ \hline

Pinaya~\cite{pinaya2023generative} &     Generative (multiple)   &  Multiple  &  Multiple   &  MONAI: generative AI library for medical imaging         \\ \hline

Gosselin~\cite{gosselinDevelopmentNewGeneration2014} &    Patient-Based Model    & MRI & Whole Body & Virtual Population VIP3.0: High-resolution models created from patient MRI                           \\ \hline

Solomon~\cite{solomon2014generic} &     Patient-Based Model    &  CT  &  Multiple    &   A simulation model for lung, liver and renal lesions                           \\ \hline

Tomic~\cite{tomic2022development} &     Patient-Based Model    &  Any  &  Breast   &    A growing tumor model for breast cancer analysis                            \\ \hline

Al Khalil~\cite{al2020heterogeneous} &     Patient-Based Model    &  MRI  &  Cardiac  &   A set of simulated models for cardiac segmentation analysis                               \\ \hline

Segars~\cite{segarsDevelopmentPopulation4D2015} &     Patient-Based Model    &  PET-CT  & Whole Body &    Extended set of pediatric XCAT phantoms                              \\ \hline

Hoe~\cite{hoe2006simulation} &     Patient-Based Model    &  Any  &  Liver    &    CT-derived pediatric liver lesion model                       \\ \hline 

Shaheen~\cite{Shaheen2014} &     Patient-Based Model    &  Multiple  &  Breast    &    An MRI-derived lesion model for DM and DBT analysis                       \\ \hline 

Sarno~\cite{Sarno2021_digitaldataset} &     Patient-Based Model    &  CT  &  Breast  &   A set of digital CT breast phantoms designed for virtual clinical trials                        \\ \hline

Sauer~\cite{sauer2023development} &     Patient-Based Model    &  CT  & Lung  &  A growing lesion model for lung analysis applications                              \\ \hline

Graff~\cite{graffNewOpensourceMultimodality2016} &     KB Model    &  Multiple  & Breast   &    A high-resolution breast phantom                             \\ \hline

de Sisternes~\cite{de2015computational} &     KB Model    & Multiple   & Breast        &  A growing breast lesion model                 \\ \hline

Sengupta~\cite{sengupta2021computational} &     KB Model    & Multiple   & Breast          &    A growing breast lesion model                      \\ \hline

Sizikova~\cite{sizikova2023knowledge} &     KB Model    &  DM  &  Breast         &   A simulated image dataset for comparative analysis of mammography AI                      \\ \hline

Abadi~\cite{abadi2018dukesim} &     Imaging Simulator    &  CT  & Any  &  DukeSim: a scanner-specific CT simulation framework                               \\ \hline 

Wu~\cite{wu2022xcist} &     Imaging Simulator    &  Multiple  &  Any         &  XCIST: An X-ray/CT simulation framework                      \\ \hline 

Badal~\cite{badal2009accelerating} &     Imaging Simulator    & Multiple   & Multiple      &   Acceleration of Monte Carlo simulations in imaging using a GPU                        \\ \hline 

Badal~\cite{badal2021mammography} &     Imaging Simulator    &  Multiple   &  Breast      &  MC-GPU: DM and DBT breast imaging simulation framework                       \\ \hline

Sarrut~\cite{sarrut2022opengate} &     Imaging Simulator    & Multiple   & Multiple    &   OpenGATE: an open-source Monte Carlo toolkit for medical physics          \\ \hline

Liu~\cite{liu2016fast} &     Imaging Simulator    &  MRI  &  Multiple         &  MRiLab: MRI simulation framework                      \\ \hline 

Unberath~\cite{unberath2018deepdrr} &     Imaging Simulator    & X-ray   &  Multiple   &  DeepDRR: simulation of X-ray from CT     \\ \hline

Jensen~\cite{jensen1997field} &     Imaging Simulator    & Ultrasound   & Multiple       &  Field: ultrasound simulation framework  \\ \hline 

Maier~\cite{maier2018deep} &     Hybrid, Physics-Informed    & CT   &  Any   &  Deep scatter estimation (CT) for real time X-ray scatter in cone-beam CT                        \\ \hline 

Horger~\cite{horger2018towards} &     Hybrid, Physics-Informed    & Any   &  Any   &  Efficient NN-based noise sampling for physics simulations                        \\ \hline 

Maier~\cite{maier2018precision} &     Hybrid, Physics-Informed    & X-ray   &  Any   &  Precision learning: incorporating priors into data-driven material decomposition \\ \hline 

\hline

\multicolumn{5}{|c|}{}  \\ 
\multicolumn{5}{|c|}{\textbf{Applications}}  \\ 
\multicolumn{5}{|c|}{}  \\ 
\hline

Teixeira~\cite{teixeira2018generating} &     GAN    & X-ray   &  Chest and Torso & X-ray synthesis from surface geometry  \\ \hline 

Frid-Adar~\cite{frid2018gan} &     GAN    & CT   &  Liver   & Generative lesion synthesis  \\ \hline 

Azizmohammadi~\cite{azizmohammadi2022generative} &     GAN    & X-ray   &  Heart & Generative learning to predict angiography frames  \\ \hline 

Ben-Cohen~\cite{ben2019cross} &     GAN    & Multiple   &  Liver   & CT to PET cross-modal synthesis for improved lesion detection  \\ \hline 

Mahmood~\cite{mahmood2018unsupervised} &   GAN      & Endoscopy   &  Chest and Torso &  Reverse domain adaptation to match real and synthetic images  \\ \hline

Shin~\cite{shinMedicalImageSynthesis2018} &     GAN    & MRI   &  Brain & Synthesis of abnormal images for improved tumor segmentation \\ \hline 

Lewis~\cite{lewis2021improving} &     GAN    & Multiple   &  Chest and Torso   & CT generation from X-ray for low-resource environments  \\ \hline 

Korkinof~\cite{korkinof2019mammogan} &     GAN    & DM   &  Breast & MammoGAN: generative synthesis of mammograms  \\ \hline 

Sun~\cite{sun2021deep} &     GAN    &  MRI  &  Vertebrae   &  Private data sharing of medical images                           \\ \hline

Thambawita~\cite{thambawita2022singan} &     GAN    &  Endoscopy  &  Gastrointestinal (GI)   &  SinGAN-Seg: synthetic data generation for improving polyp segmentation                           \\ \hline

Zunair~\cite{zunair2021synthesis} &     GAN    &  X-ray  &  Chest and Torso   &  Synthetic COVID-19 chest X-ray dataset created using CycleGAN         \\ \hline

Salem~\cite{salem2019multiple} &   Generative      & MRI   &  Brain &  Multiple Sclerosis (MS) lesion synthesis  \\ \hline 

Prados~\cite{prados2016fully} &   Generative      & MRI   &  Brain & Pathology inpainting in MRI for Multiple Sclerosis (MS) applications  \\ \hline 

Konukoglu~\cite{konukoglu2013example} &     Generative    & MRI   &  Any & MRI artifact reduction  \\ \hline 

Li~\cite{li2022cross} &     VAE    & MRI   &  Lung & Cross-modality synthesis for improved tumor segmentation \\ \hline 

Chambon~\cite{chambon2022roentgen} &     DDPM    & X-ray   &  Chest and Torso & RoentGen: text-to-image synthesis of X-rays  \\ \hline 

Stojanovski~\cite{stojanovski2023echo} &     DDPM    &  Ultrasound  &  Cardiac   &  Generation of synthetic ultrasound images for improved segmentation         \\ \hline

Ali~\cite{ali2022spot} &     DDPM    &  Multiple  &  Chest and Torso   &  Spot the fake lungs: generation of synthetic X-ray and CT lung images         \\ \hline

Gao~\cite{gao2022synthex} &     Patient-Based Model    & CT   &  Multiple & SyntheX: CT to X-ray simulation using DeepDRR  \\ \hline 

Xanthis~\cite{xanthis2021simulator} &     Patient-Based Model    & MRI   &  Heart & Simulation of synthetic images via XCAT for improved segmentation  \\ \hline 

de Dumast~\cite{de2022synthetic} &     Patient-Based Model    & MRI   &  Fetal & Simulation of fetal brain MRI from a phantom for domain adaptation  \\ \hline 

Pezeshk~\cite{pezeshk2015seamless} &     Patient-Based Model    & CT   &  Lung   & Synthetic data generation using lesion insertion  \\ \hline 

Brumer~\cite{brumer2022synthetic} &     Patient-Based Model    &  MRI  &  Kidney   &  Arterial spin labeling (ASL) analysis using synthetic data generated from XCAT          \\ \hline

Badano~\cite{Badano2018victre} &     KB Model    &  Multiple  &  Breast   &  VICTRE: virtual clinical trial for comparing DM and DBT                           \\ \hline

Kadia~\cite{kadia2022lesion} &     KB Model    & CT   &  Any   & A lung lesion model for improvement of segmentation performance \\ \hline 

Gong~\cite{gong2006computer} &     KB Model    & Multiple   &  Breast   & Simulations-based comparison of DM, DBT, and cone-beam CT imaging  \\ \hline 

Nelson~\cite{nelsonPediatricSpecificEvaluationsDeep2023} &     KB Model    & CT   &  Multiple   & Assessment of DL-based CT denoising using computer simulations  \\ \hline 

Li~\cite{li2009three} &     KB Model    &  CT  &  Lung   &  A lung nodule model for pediatric CT \\ \hline

Cha~\cite{cha2020evaluation} &     Any    & DM   &  Breast & Analysis of data augmentation via synthetic data for breast mass detection \\ \hline 

\multicolumn{5}{|c|}{}  \\ 
\multicolumn{5}{|c|}{\textbf{Surveys and Overviews}}  \\ 
\multicolumn{5}{|c|}{}  \\ 
\hline

Kazerouni~\cite{kazerouni2023diffusion}       &     \multicolumn{3}{|l|}{Survey/Overview}      &    Survey of DDPM models in medical imaging                          \\ \hline

Singh~\cite{singh2021medical}              &     \multicolumn{3}{|l|}{Survey/Overview}        &    Survey of GAN models in medical imaging                       \\ \hline

Pesapane~\cite{pesapane2022digital}  &   \multicolumn{3}{|l|}{Survey/Overview}   & Discussion of digital twins in radiology   \\ \hline

Badano~\cite{badano2023stochastic} &   \multicolumn{3}{|l|}{Survey/Overview}  &  Survey of techniques for generating synthetic data models  \\ \hline 

Xu~\cite{xuExponentialGrowthComputational2014} &   \multicolumn{3}{|l|}{Survey/Overview} &  Survey of computational phantom use for radiation dose quantification  \\ \hline 

Segars~\cite{segarsApplication4DXCAT2018}  &   \multicolumn{3}{|l|}{Survey/Overview}   & Survey of 4-D XCAT applications   \\ \hline 

Kainz~\cite{kainzAdvancesComputationalHuman2019}                                 &     \multicolumn{3}{|l|}{Survey/Overview}     &     Survey of computational human phantoms                             \\ \hline

Killeen~\cite{killeen2023silico} &     \multicolumn{3}{|l|}{Survey/Overview}      &    Survey of in silico simulation for minimally invasive surgery  applications                         \\ \hline 

Rodero~\cite{rodero2023systematic} &   \multicolumn{3}{|l|}{Survey/Overview} &  Survey of in silico clinical trials for cardiac applications  \\ \hline 

Guan~\cite{guan2021domain}  &   \multicolumn{3}{|l|}{Survey/Overview}   & Survey of domain adaptation for medical image analysis   \\ \hline 

Candemir~\cite{candemir2021training} &   \multicolumn{3}{|l|}{Survey/Overview} &  Survey of techniques for training radiologic deep learning models in data-limited scenarios  \\ \hline 

Boulanger~\cite{boulanger2021deep} &     \multicolumn{3}{|l|}{Survey/Overview}   & Survey of data-driven methods for CT synthesis from MRI  \\ \hline 

Edmund~\cite{edmund2017review}  &   \multicolumn{3}{|l|}{Survey/Overview}   & Survey of CT substitution methods in MRI-only imaging  \\ \hline 

DuMont Sch{\"u}tte~\cite{dumont2021overcoming}  &   \multicolumn{3}{|l|}{Survey/Overview}   & Benchmark of GAN techniques on chest X-ray and brain CT with a study of data sharing applications   \\ \hline 

Castiglioni~\cite{castiglioni2021ai}  &   \multicolumn{3}{|l|}{Survey/Overview}   & Survey of medical imaging AI systems used as clinical decision support tools   \\ \hline

Kelkar~\cite{Kelkar2023}          &     \multicolumn{3}{|l|}{Survey/Overview}        &   Overview of GAN assessment in medical imaging                       \\ \hline

Dar~\cite{dar2023investigating}  &   \multicolumn{3}{|l|}{Survey/Overview}   & Assessment of data memorization in medical DDPM models  \\ \hline 

Shorten~\cite{shorten2019survey} &   \multicolumn{3}{|l|}{Survey/Overview} &  Survey of image data augmentation in deep learning  \\ \hline 

\end{tabular}}
}
\caption{\small Types of models, application modalities and anatomies analyzed using synthetic data.}
\label{table:survey}
\end{table}

\section{Techniques for Synthetic Data Generation}
Techniques for synthetic imaging data generation can be broadly grouped into three categories: statistical generative modeling, physics-based modeling, and hybrid, physics-informed modeling. A summary of popular models, applicable imaging modalities anatomies can be found in Table~\ref{table:survey}.  

\subsection{Statistical Generative Models} 
Generative models learn to synthesize outputs (images) that capture patterns and structures observed from existing patient images~\cite{singh2021medical} by processing the distribution of pixel intensities. Most recent models are based on various neural network architectures developed in the ML community. 

\textbf{Generative Adversarial Networks (GANs).} The key idea behind GANs, a popular type of generative model, involves two competing networks~\cite{goodfellow2020generative}: the first network (generator) aims to synthesize data that resembles the distribution of real data while the second network (discriminator) aims to differentiate the synthetic data from the real data. {\color{black} The GAN training process is adversarial and approximately solves a min-max optimization problem, with the objective of creating new data that matches the statistical distribution of training data.} GANs have been used for generating synthetic training images~\cite{singh2021medical}, creating annotations~\cite{kossen2022toward}, cross-domain~\cite{jiang2018tumor} and pseudo-healthy synthesis~\cite{xia2020pseudo}. {\color{black} Extensions of GANs include CycleGAN~\cite{zhu2017unpaired}, which enables image domain transformation without the need for paired data, and AmbientGAN~\cite{bora2018ambientgan}, which learns implicit generative models from lossy measurements of the distribution of interest. Both variants have found numerous applications in medical imaging~\cite{zhou2022learning}.}

\textbf{Normalizing Flow (NF).} Normalizing flows are part of the generative model family that learn an invertible transformation, typically represented by a neural network, from a well-understood base distribution (e.g., multivariate Gaussian) to a complex data distribution~\cite{kobyzev2020normalizing}. This base distribution serves as the starting point (``prior'') from {\color{black} which data is generated}. NF learns a series of transformations to morph this base distribution to the target data distribution, enabling the generation of synthetic data that mimics the original. As compared to GANs, NFs offer an opportunity for a more profound interaction with the inherent data properties. While NFs do not inherently model physical properties of the data, their ability to provide exact likelihood evaluation allows them to better capture these properties if they significantly influence the data distribution. Nevertheless, incorporating domain-specific knowledge or physical laws directly into the flow structure is still an active area of research~\cite{kobyzev2020normalizing}.
In radiology, NFs have recently gained some attention in applications such as {\color{black} image reconstruction~\cite{denker2020conditional}} and data augmentation~\cite{hajij2022normalizing}. 

\textbf{Variational Autoencoders (VAEs).}  VAEs leverage the principles of autoencoding and variational inference~\cite{kingma2013auto}, and consist of two components: an encoder network and a decoder network. The encoder transforms the input data into a specific distribution in the latent space. The decoder samples points from this latent distribution and attempts to reconstruct the original data. Through this process, VAEs can learn a stochastic, continuous {\color{black} bidirectional} mapping between the data and latent space. {\color{black} When only a limited number of training examples are available, combining variational inference with GANs may help avoid mode collapse, i.e., generation of uniform or blurry examples~\cite{ahmad2022brain}.}

\textbf{Denoising Diffusion Probabilistic Models (DDPMs).} DDPMs are a type of generative model that {\color{black} represents image formation} as a diffusion process. This process starts with the actual data and gradually adds noise until a simple noise distribution is reached~\cite{ho2020denoising}. To generate new data, the procedure is reversed by taking a sample from the simple noise distribution and iteratively applying a learned denoising operation, until the original data distribution is recovered. Here, noise operations are typically parametrized by a deep neural network, allowing the model to learn complex transformations between the noise and data distributions. By using a noisy and stochastic transition process, DDPMs are able to model a wide variety of data distributions. Compared to GANs and VAEs, DDPMs may be easier to train and have a faster inference time~\cite{kazerouni2023diffusion}.

\subsection{Physical Modeling}
Synthetic data generation using physical modeling typically includes two components~\cite{badano2023stochastic}: a digital model representing a patient or patient populations, and a digital model of an acquisition device (imaging system). 

\subsubsection{Digital Human Models}
Digital human models for computational simulations have been developed extensively over the past decades for different applications, particularly radiation dosimetry~\cite{xuExponentialGrowthComputational2014}. Recent research has focused on the development of models with increased spatial resolution and anatomical realism. The level of detail and anatomic diversity in these models depend on the method of generation and the range of anatomy covered (whole body or specific regions). 
The majority of digital human models are derived from detailed segmentations of tomographic images of patients~\cite{segarsApplication4DXCAT2018}. The voxelized organs resulting from the segmentation process can be  converted to surface mesh models to allow modifications and repositioning.
Each organ is then assigned appropriate material properties depending on the intended use of the model. An early example is the Virtual Population VIP3.0~\cite{gosselinDevelopmentNewGeneration2014}, a collection of digital human models developed for electromagnetic (EM) exposure evaluations.
{\color{black} Another popular digital human model, }the Extended Cardiac Torso (XCAT) phantom~\cite{segarsApplication4DXCAT2018}, used a few reference surface phantoms and registered them to patient images to create large cohorts of digital models. The XCAT incorporates respiratory and cardiac motion, and has sufficient resolution to be used in imaging. Detailed digital human models have been used extensively in a range of applications~\cite{kainzAdvancesComputationalHuman2019}, ranging from developing image processing and reconstruction methods to motion compensation.
Anatomic models of specific parts of the body are also commonly used, particularly for breast imaging applications. For example, a procedurally-generated stochastic breast model including a skin layer, blood vessels, glandular ducts, fat and other components was created by Graff et al.~\cite{graffNewOpensourceMultimodality2016}, and used in the evaluation of full-field digital mammography and tomosynthesis in the Virtual Imaging Clinical Trial for Regulatory Evaluation (VICTRE) project~\cite{Badano2018victre}.

\subsubsection{Digital Acquisition Device Models}
Radiological images can be reliably simulated in silico because the physical processes underlying the generation, propagation and detection of radiation (from optical light to gamma rays) are well understood. Physics-based digital replicas of radiation sources and detectors, coupled with realistic transport of radiation through digital phantoms, are used to create synthetic images that reproduce the features of images acquired with physical devices. The required accuracy of the image generation process depends on the context of use (COU) of the images. Typically, more realistic images can be generated by implementing more sophisticated physics models, at the expense of increasing the computational complexity. As an example, x-ray projections of digital phantoms can be efficiently simulated using Siddon's ray-tracing algorithm, which models x-rays as straight lines from the source to the center of each pixel. However, if the pixel noise statistics or the contribution from scattered radiation are relevant to the context of use, more sophisticated Monte Carlo (MC) methods that track the interactions of individual x-rays might be necessary. 

Numerous software packages have been developed to simulate different imaging modalities. For example, DukeSim~\cite{abadi2018dukesim} and XCIST~\cite{wu2022xcist} simulate commercial computed tomography (CT) scanners using a combination of ray-tracing and MC methods. MC-GPU~\cite{badal2009accelerating,badal2021mammography} implements a GPU-accelerated MC code for cone-beam computed tomography (CT), mammography and tomosynthesis (as shown in Fig. 1). Nuclear medicine applications can be simulated with the MC tools from the OpenGATE Collaboration~\cite{sarrut2022opengate}. 
Simulation packages for imaging modalities not using ionizing radiation such as magnetic resonance imaging (MRI)~\cite{liu2016fast} and {\color{black} ultrasound~\cite{jensen1997field}} are also available. \textcolor {black} {A comprehensive review of various simulation frameworks for visible light (endoscopic), ultrasound and x-ray imaging, as well as their applications in intelligent surgical systems is given by \cite{killeen2023silico}.}   

\subsection{Hybrid, Physics-Informed Models}
Although recent advances in parallel computing, including graphical processing units (GPU), have allowed for complex physics-based simulations, such simulations may still often be prohibitive due to the computational {\color{black} overhead}. Hybrid, physics-informed models address this concern by accelerating select components of synthetic data generation with deep learning. Alternatively, physics-informed neural networks embed physical constraints to create more realistic outputs or reduce the amount of training samples needed to learn a task. For example, deepDRR~\cite{unberath2018deepdrr} speeds up generation of fluoroscopy and digital radiology from computed tomography (CT) scans by performing ML for scatter estimation and material decomposition, while retaining an analytic approach for other pipeline components. There are several other examples. \cite{maier2018deep} proposed a deep scatter estimation (DSE) technique that is within 2\% of traditional Monte Carlo simulations used for cone beam CT acquisition. In fact, a neural network (NN) can learn to sample from a given probability density function (PDF) with high sampling efficiency~\cite{horger2018towards}, making it useful for noise modeling in physical simulations~\cite{maier2019gentle}. {\color{black} Finally, when known operators are combined together with NNs to inform the latter about known prior information during the training and inference, a NN may require less training data, training iterations, or achieve better performance levels~\cite{maier2018precision}.} 

\subsection{Synthesizing Disease Models}
The lack of well-curated and labeled data is particularly acute for diseased cases. Synthetic examples generated using generative modelling have been explored for creating various types of lesions~\cite{kadia2022lesion,salem2019multiple,liu2022free}. Alternatively, lesions could also be synthetically in-painted (i.e., removed) to reduce impact on image processing tasks such as registration or segmentation~\cite{prados2016fully}. {In silico, knowledge-based models of disease have been developed for various organs~\cite{de2015computational,solomon2014generic,rodero2023systematic}.} An important consideration for lesion models is their growth pattern~\cite{sengupta2021computational,tomic2022development}, since lesion presentation may be affected by properties of the surrounding tissue.

\subsection{Limitations of Data Generation Techniques}
Statistical generative models are typically trained using images (e.g., collections of x-rays) and are able to rapidly generate examples from the learnt generative distributions. However, they may not learn appropriate physical constraints or causal links between attributes and physical findings, and thus often suffer from generating hallucinated findings or unrealistic anatomy.  On the other hand, physics-based approaches are grounded in physiology naturally embedded in the digital human model, and are able to generate high-quality and fully-detailed outputs controlled by the input parametrization. These approaches, however, may require more time-consuming and computationally intensive simulations. In addition, physical modelling approaches are constrained by the variability of the parameter space of the digital human model and acquisition system, but the complexity of the model can be adjusted based on the task of interest. Hybrid, physics-informed models are typically designed to accelerate components of physics-based approaches using neural networks, which may result in loss of realism, limited variability or constrained generalization~\cite{unberath2018deepdrr}. {\color{black} The \textit{generative learning trilemma} states that current data generation approaches cannot generate high-quality, diverse samples fast enough ~\cite{xiao2021tackling}.} On the other hand, generative models and mechanistic physical models may be complementary~\cite{an2022integration}. 

\textbf{Patient-Derived Models.}
All models (whether generative or physics-based), that are created solely based on a fixed set of patients are limited to properties (e.g., presence of disease) observed during training, rather than full object properties that may characterize the population of interest. {\color{black} For example, patients with advanced breast cancer may be not captured by a patient-derived model that did not include such patients in the training set.} Such properties are better captured by knowledge-based models, derived from physical or biological measurements{\color{black}, to the extent that the knowledge is representative of the patient population}. In addition, patient-derived models may be constrained by the quality {\color{black} and resolution} of the training data, including noise, artifacts, contrast constraints, and missing data.

\textbf{Null Space.} Image-based methods, whether parametric or generative, are limited by the existence of a null space, which results from mapping a continuous object to a discrete image by an acquisition system, resulting in an unavoidable loss of information~\cite{tam2012null}. This limitation can be addressed by either learning from object models (typically via physics-based simulations) or by modifying the generative model training process to capture image degradation during training~\cite{zhou2022learning}. 

\textbf{Realism.} A key concern in the use of any synthetic data is its realism, i.e., the size of the distribution gap between real and synthetic examples, particularly in areas that affect device performance. {\color{black} Prior to integration of real and synthetic datasets, some pre-processing methods can be implemented to reduce the distribution gap, either using engineered features or learnt image transformations~\cite{guan2021domain,gao2022synthex,zhu2017unpaired}. However, the problem of mitigating the `synth2real' gap still remains a hurdle.  Also, one of the drawbacks of using statistical generative models for data augmentation is that the supplementary generated examples may not extend beyond the training distribution of the model.}

\section{Applications}

\subsection{Algorithm Development and Training} Synthetic examples have been widely used as a source of training data, either on their own or combined with real patient images. This approach has been well-explored across many types of radiological imaging~\cite{candemir2021training}. For instance, \cite{teixeira2018generating} showed that augmenting limited patient x-rays with synthetic images reduced marker localization error. \cite{cha2020evaluation} demonstrated that the addition of synthetic mammograms generated using in silico imaging improved performance according to breast mass detection free-response receiver operating characteristic (FROC) as compared to results from patient data alone. Several studies have used GANs for data augmentation and improved the performance of their algorithms, as seen in liver lesion classification on CT images~\cite{frid2018gan}, brain segmentation on CT and MRI images~\cite{bowles2018gan}. {\color{black} Synthetic images can address class imbalance concerns, but only if the synthetic images deviate sufficiently from the existing patient data~\cite{candemir2021training}.}

\textbf{Image Reconstruction and Cross-modality Synthesis.} \label{sec:cross_modal_rec}
There has been a number of works that aim to predict one modality (e.g., CT) from another (e.g., x-ray) for improving image quality and decreasing number of artifacts~\cite{konukoglu2013example}, reducing radiation exposure~\cite{azizmohammadi2022generative}, and {\color{black} improving prediction accuracy (e.g., lesion detection~\cite{ben2019cross})}. CT prediction from MRI has been particularly well-explored~\cite{boulanger2021deep}, as tissue electron density information from CT is needed for radiotherapy planning~\cite{edmund2017review}. 

\textbf{Source of Annotations.} {\color{black}A significant advantage of synthetically generated examples is that they can be generated to include pixel-level annotations needed for training AI algorithms, thus reducing the annotation burden while retaining or even improving accuracy~\cite{al2020heterogeneous,gao2022synthex}, since pixel-level labels are particularly challenging to annotate. Segmentation supervision for images can be obtained either using deep conditional generation~\cite{jiang2020covid} (i.e., generating images conditionally on an input segmentation mask) or using simulation~\cite{xanthis2021simulator,mahmood2018unsupervised} (where segmentation truth is obtained from a digital model).}

\subsection{Algorithm Testing} {\color{black} Synthetic data can be used for generating standardized testing examples that would be otherwise too difficult to acquire from patient images~\cite{gong2006computer,segarsDevelopmentPopulation4D2015}. When a synthetic dataset is used for testing, it is particularly important to ensure that this dataset is representative of the intended patient population in order for performance estimates to be accurate. Thus, compared to the scenario where synthetic data is used for training, the evaluation requirements for synthetic testing data are more stringent. } 

Sizikova et al. \cite{sizikova2023knowledge} introduced the idea of using synthetic images for comparative performance testing in medical imaging, where AI is evaluated on known trends with respect to physical properties (e.g., mass size). For this application, physics-based synthetic simulations are particularly useful since they can be used to easily re-generate examples with modifications to physical properties (e.g., size or radiation dose), while obtaining similar patient examples may not be practically possible. An emerging application of synthetic data is within in silico clinical trials, where results from computer simulations are used in development or regulatory evaluation of a medicinal product, device, or intervention\textcolor{black}{~\cite{viceconti2016silico, rodero2023systematic}}. Here, synthetic data complements patient data for evaluation of novel treatment methodologies or medical devices. \cite{Badano2018victre} has shown that an in silico clinical trial comparing digital mammography (DM) and digital breast tomosynthesis (DBT) imaging modalities replicated the results of an in situ (non synthetic) clinical study involving hundreds of enrolled women. As discussed in \cite{badano2021silico}, in silico trials are not identical to their in situ counterparts, and could provide evidence not found in traditional clinical trials~\cite{faris2017fda}. 

\subsection{Patient Privacy Preservation} Synthetic data can act as an anonymization tool to protect patient characteristics while sharing data. For instance, a recent study~\cite{dumont2021overcoming} has evaluated the quality of GAN-generated synthetic chest radiographs as an alternative to sharing patient chest radiographs and brain CT, and showed that NN performance matched closely when trained on either synthetic or real examples, but suffered when a larger number of classes (labels) was considered. {\color{black} However, the risk of generative models inadvertently memorizing specific data points, thereby compromising patient privacy, cannot be ignored~\cite{carlini2023extracting}.  We refer the reader to \cite{giuffre2023harnessing} for a discussion of synthetic healthcare data privacy and associated risk mitigation measures. Finally, a recent set of recommendations for utilizing and evaluating differential privacy (AI) published by the National Institute of Standards and Technology (NIST) can be found in \cite{near2023guidelines}.}

\subsection{Addressing Bias and Other Limitations of Patient Datasets}

\subsubsection{Class Imbalance and Modality Availability} Many datasets are prone to data imbalance, i.e., an uneven data distribution across classes, due to, for instance, the secondary use of data~\cite{tucker2020generating}.  {\color{black} A popular technique to address this issue in imaging studies is the use of resampling techniques that synthetically resize training datasets to obtain more balanced distributions~\cite{castiglioni2021ai}. Algorithmic fairness approaches~\cite{hardt2016equality} may be used to balance out uneven distributions in available patient datasets. \cite{ktena2023generative,khosravi2023few} demonstrated the benefits of synthetic radiologic data created using generative models conditioned on various input attributes, such as examples with limited annotations or less frequent categories, to address class imbalances arising from existing data. As discussed in Section~\ref{sec:cross_modal_rec}, synthetic data can also be used to impute missing information. }

\subsubsection{Enrichment of Underrepresented Populations} An attractive feature of synthetic data is that it can be used to generate examples from known and underrepresented populations, such as patients with rare diseases and protected populations. For instance, \cite{shinMedicalImageSynthesis2018} reported the creation of synthetic training data by inserting rare abnormal tumors into MRIs and demonstrated that this process improved model performance on patient examples. in another work, \cite{lewis2021improving} investigated TB classification using synthetically generated CT over patient X-ray alone, and discussed the potential applications of synthetic data in supplementing costly imaging procedures for resource-poor communities.

Protected populations are also notoriously difficult to obtain data points from, and are a potential candidate for synthetic data use. For example, pediatric patients represent 20\% of the US population, but make up only 5\% of imaging studies ~\cite{smith-bindmanTrendsUseMedical2019}. Pediatric radiology datasets are particularly difficult to acquire due {\color{black} to a} lack of domain specialist annotators, lower study numbers, added safety concerns and regulatory requirements, all of which contribute to a lack of AI applications for these patients~\cite{sammer2023use}. \cite{nelsonPediatricSpecificEvaluationsDeep2023} simulated pediatric-size phantoms to evaluate AI denoising algorithms in newborn to adolescent sizes. \cite{hoe2006simulation} demonstrated that synthetically generated pediatric liver CT images with in-painted lesions were indistinguishable to real counterparts when read by radiologists. \cite{li2009three} showed that CT images of synthetic and patient lung nodules in pediatric patients were perceptually indistinguishable. {\color{black} Finally, \cite{de2022synthetic} explored applications of synthetic data to AI-based segmentation of brain tissues in fetal MRI. 

\textbf{Considerations.} A key challenge in synthetic data use for underrepresented populations is that it is inherently hard to find samples to build a robust training dataset for the data generation model to ensure that it does not perpetuate existing biases~\cite{luccioni2023stable}. While approaches such as class-specific few-shot learning~\cite{ruiz2023dreambooth,chambon2022roentgen} may mitigate the issue, under such conditions, physical modeling, which typically requires fewer parameters, may be advantageous. In either case, attention must be given not to perpetuate existing biases present in the data or the knowledge model. }

\section{Data Assessment Metrics}
\textbf{Fidelity and Utility.} Evaluating synthetic radiological data is important to ensure that the generated data can serve its intended purpose. Synthetic data is often assessed in terms of its \textit{fidelity}, i.e., whether it captures statistical inter-relationships of patient datasets, or its \textit{utility}, i.e., whether it achieves similar results (e.g., downstream task performance) as patient data. A high-fidelity dataset therefore should have high utility~\cite{myles2023potential}, however, high utility may not be necessary for applications such as an understanding of relative trends. {\color{black}Fidelity metrics (e.g., Frechet Inception Distance (FID))  may capture summary statistics, single or pairwise distributional patterns, more complex interrelationships between variables in the synthetic and/or patient data points and or consistency with clinical domain expertise~\cite{Kelkar2023}.} As the number of data dimensions increase, measuring fidelity becomes increasingly complex due to the exponentially increasing number of interrelationships~\cite{platzer2021holdout}. 

\textbf{Types of Utility Metrics.} As discussed in \cite{myles2023potential}, utility metrics measure distance between patient and synthetic data and could be grouped into work-aware evaluations (metrics that compare real and synthetic data performance in tasks of interest), generic utility measures (metrics that compare general distance metrics between real and synthetic data), and subjective assessments (metrics that compare perceptual quality of synthetic and patient data based on domain expert assessment in a user study). How to evaluate synthetic radiological data is an on-going and open research problem.  \cite{korkinof2019mammogan} compared synthetic and real examples using distributions of top five moments. Work-aware evaluations (alternatively, task-based performance) is another popular method of evaluating synthetic data~\cite{frid2018gan}, particularly when the underlying real data distribution is unknown. Objective assessment of image quality in medical applications has been extensively studied~\cite{barrett2013foundations}. While current metrics provide some level of assessment for synthetic radiological data, they have limitations~\cite{alaa2022faithful,platzer2021holdout}, and may overlook issues such as hallucinations and memorization in synthetic radiological data~\cite{bhadra2021hallucinations}. 

\textbf{Subjective Assessment.} Subjective assessment metrics~\cite{chow2016review} typically include user studies or visual clinical assessments with reader studies. \textcolor{black}{
One well-known example is the visual Turing test (sometimes known as a ``fool the human'' study design), a commonplace approach to evaluate the realism of synthetic images~\cite{pezeshk2015seamless} that derives from Turing's work on machine intelligence~\cite{TTObjections}. In a Turing test, radiologists are tasked with distinguishing between synthetic and real images. Several studies found that the performance of radiologists is at almost guessing levels~\cite{de2015computational,pezeshk2015seamless,solomon2014generic}, since a visual Turing test includes subjective and highly noisy nature of the interrogator~\cite{TTObjections}. \cite{Shaheen2014} described a study where five radiologists were asked to evaluate mammography and breast tomosynthesis (DBT) images, comparing mass appearance in patient images and in images where MRI-derived, spiculated lesions were artificially inserted. Results showed that humans were only marginally capable of distinguishing synthetic from natural masses with an area-under-the-ROC curve averaged across observers of 0.54 (95\% confidence interval [0.50, 0.66]) for the 2D study and 0.67 (95\% confidence interval [0.55, 0.79]) for tomosynthesis. It is important to note that the result of the Turing test (whether human readers can correctly identify synthetic samples) may or may not be relevant for determining whether computational models and simulation tools are useful for evaluating new imaging systems~\cite{realism2017}.
}

\textbf{Future Directions} 
A number of existing metrics for evaluating synthetic data already exist today~\cite{alaa2022faithful}. One future direction would involves establishing a standardized and comprehensive framework that can provide a holistic evaluation of synthetic data, ensuring that metrics are tailored to the application and offering a multifaceted view of performance, verifying that the synthetic data meets its intended purpose. 

\section{Challenges for the Use of Synthetic Data}
{\color{black} As in other fields, applications of synthetic data in radiological imaging suffer from data complexity (multi-scale models spanning several scientific disciplines), disclosure limitations (no robust platform to develop and disseminate models), data privacy and data ownership concerns. Below, we discuss some existing challenges associated with practical use and evaluation of synthetic radiological data.

\subsection{Scientific Challenges}
To ensure the safety and effectiveness of new biomedical technologies developed or evaluated with synthetic data, additional research is needed to better understand the uncertainty and bias of synthetic data generation approaches. An open area of research is the development of metrics for characterizing individual or population representation in a synthetic dataset and for evaluating the reliability of algorithm performance (e.g., does the algorithm performance reported on a synthetic dataset match the performance on a real patient dataset).} In addition, techniques to ensure unbiased outcomes from utilizing synthetic data need to be developed, in particular, to ensure that using synthetic data does not lead to shortcuts as compared to real examples. Moreover, statistical methods that optimize the incorporation of synthetic data into small patient test datasets need to be investigated. Finally, there is a need for methods and tools to quantify and improve the generalizability of the synthetic data when replicating different scenarios in terms of the clinical settings, population characteristics, and imaging device properties. In this context, methods and techniques to safely reuse synthetic data in the development and evaluation stages of new technology, as well as in  post-market “in-the-wild” monitoring programs are needed including sound methods for data reuse and for practical and efficient data sequestration strategies.

\subsection{Evaluation Challenges} 

The key challenge of using synthetic data in the context of medical device evaluation concerns validation requirements that should be sufficiently strict to support the data usage. Synthetic data may be used in multiple ways within the context of regulatory evaluations with different evidentiary requirements. Regardless of the specific application, evidence must exist to show that synthetic data within a regulatory submission can be sufficiently relied upon to support the claims made. Depending on how the synthetic data is used, this may include synthetic examples with patient data, cross-validation of synthetic data using different data generation techniques, or distribution gap analysis between patient and synthetic data. The greater the prominence of synthetic data in the submission, the higher the evidentiary requirements. As an example, a computer-aided diagnostic device may use a data augmentation strategy incorporating synthetic examples into its training data. Validation of such supplementary data may include demonstrating that the synthetic data distribution follows a similar distribution in specific features as with real data, and that its use improves performance of the studied algorithm in real data. Primary research supporting the safety and effectiveness of the device would continue to come from a robust test set of real patients.

As all models carry inherent assumptions, ensuring congruence between the use of the synthetic data in technology assessment and the original purpose of the data generation model is essential. The US Food and Drug Administration (FDA) applies the term \textit{``Context of Use (COU)''} to describe the way the synthetic data/algorithm/model is to be used. For a radiological device reliant on an AI algorithm, the context of use is the specific role and scope of the device, a detailed description of what will be modeled and how the device outputs will be used to answer specific questions of interest\footnote{See FDA guidance titled ``\textit{Assessing the Credibility of Computational Modeling and Simulation in Medical Device Submissions}'' [\url{https://www.fda.gov/media/154985/download}].}. The context of use, the influence of the model in the regulatory decision, and the consequences of  decision on patients, inform the validation and performance criteria necessary for the synthetic data to be relied upon for regulatory purposes.

\section{Summary}
Synthetic data shows great promise in advancing radiological imaging, especially AI-based technologies. Developing or evaluating these technologies with synthetic data allows conserving resources and can help to ensure  that device approvals consider the entire intended patient population. Looking forward, a recent study~\cite{wang2022development} introduced the concept of a metaverse of ``medical technology and AI'' (MeTAI) that would augment regulatory evaluation of medical devices with virtual patient and scanner models, highlighting the interplay of synthetic data and healthcare applications. Badano et al. demonstrated this proof of concept for mammographic imaging with the VICTRE trial~\cite{Badano2018victre}. Continued development and refinement of synthetic data generation techniques and applications in radiology are needed to make that future the next reality.  

\bibliographystyle{unsrt}   %
{\small
\bibliography{main}
}

\end{document}